\documentclass[aps,prd,groupedaddress, superscriptaddress, nofootinbib, reprint]{revtex4-2}

\usepackage{amsmath}
\usepackage[dvipsnames]{xcolor}
\usepackage{graphicx}
\usepackage{tikz}  
    \usetikzlibrary{positioning,decorations.pathmorphing}
\usepackage[]{hyperref}
    \hypersetup{
        colorlinks = true,
        allcolors = blue
    }

\begin{document}

\title{Schwarzschild-de Sitter spacetime in regular coordinates with cosmological time}

\author{Leonardo de Lima}
\email[]{leonardo.lima.88@edu.ufes.br}
\affiliation{Departamento de Física \& Núcleo de Astrofísica e Cosmologia (Cosmo-Ufes) \& PPGFis, Universidade Federal do Espírito Santo, Vitória, ES,  29075-910, Brazil.}
\author{Davi C. Rodrigues}
\email[]{davi.rodrigues@ufes.br}
\affiliation{Departamento de Física \& Núcleo de Astrofísica e Cosmologia (Cosmo-Ufes) \& PPGFis, Universidade Federal do Espírito Santo, Vitória, ES,  29075-910, Brazil.}
\affiliation{Centro Brasileiro de Pesquisas Físicas (CBPF), R. Xavier Sigaud 150, 22290-180, Rio de Janeiro, RJ, Brazil}

\begin{abstract}
Starting from the Einstein equations in Schwarzschild–de Sitter (SdS) spacetime and imposing Friedmann-Robertson-Walker coordinates at large distances, we find two coordinate systems with time-dependent metrics that are smooth across both the black hole and cosmological horizons. These coordinates require positive cosmological constant for them being regular, and thus they are not de Sitter extensions of the Kruskal–Szekeres or Israel coordinates. One of the coordinate systems was only found in 1999 (Abbassi coordinates), and it has led to conflicting interpretations in the literature; while the other was briefly commented and promptly dismissed as unphysical or incompatible with SdS. We derive that the second solution is equivalent to the first one, and that both are indeed equivalent descriptions of SdS spacetime. We also derive explicit coordinate transformations linking these coordinate systems to the Kottler coordinates and the maximally extended Lake–Israel coordinates. Among other applications, these results, which extend the largely used cosmological and local coordinates, should be useful for further developments on understanding the exact interplay between black holes and the cosmological background, which has been in the focus of a number of recent works.
\end{abstract}

\maketitle

\section{Introduction} \label{sec:intro}

Currently, there is renewed interest in the interplay between local and cosmological scales, in particular on possible nontrivial interactions between black holes (BHs), or BH-like objects, and cosmology (e.g., \cite{Faraoni:2007es, Croker:2021duf, Farrah:2023opk, Cadoni:2023lqe, Amendola:2023ays, Cadoni:2023lum, Cadoni:2024jxy, Faraoni:2024ghi, Croker:2024jfg, Poplawski:2024rys, Cadoni:2024rri, Calza:2024qxn}). It is not \textit{a priori} expected to find a coupling between astrophysical BHs and cosmology since, observationally, such scales are far apart by several orders of magnitude: the Schwarzschild radius of a supermassive BH is smaller than 1 pc, while the Hubble radius is larger than 10 Gpc. Moreover, the neighborhood of an astrophysical BH is not a smooth cosmological background: the accretion disk, neighboring stars, and the gravitational potential of the galaxy that hosts the BH can be more relevant than any direct cosmological influence. Thus, observationally, a spacetime that simply describes a BH in a de Sitter universe is expected to be quite a simplistic idealization.  Nevertheless, understanding such extremum idealization is a thought experiment that opens doors to understanding gravity and the universe.

A well-known general relativity (GR) solution that incorporates these extreme scales is the Kottler solution \cite{Kottler:1918cxc, 1998PhLA..245..363R, StephaniExact}, which describes the Schwarzschild-de Sitter (SdS) spacetime. This spacetime provides a clear physical scenario that GR must describe consistently. Among all exact solutions of GR representing a compact object embedded in a cosmological background, the SdS solution is the simplest.

Motivated by the approach of \cite{Cadoni:2023lqe, Cadoni:2024jxy} for deriving metric solutions for a spherical object in a cosmological background, but dropping a simplifying assumption ($\dot \alpha = 0$, as detailed in Sec.~\ref{sec:building} and Appendix \ref{app:dotAlpha}), we derive regular coordinates for SdS. The coordinates are regular in all spacetime, including the horizons, they are only singular at the center of the spherical symmetry, which is a geometrical singularity. Related to this property, and contrary to Kottler coordinates, there is no inversion between the space and time coordinates: that is, independently if a horizon is crossed, a spacial coordinate is always spacial. We stress that these coordinates are regular only for positive cosmological constant $\Lambda$, hence they are not de Sitter extensions of well-known BH regular coordinates. 

The two sets of regular coordinates that we find are related by a map that inverts their time and space natures. These solution branches are here referred to as positive and negative branches. In the limit $\Lambda \to 0$, one set can only describe the interior of the Schwarzschild BH (the negative branch), while the positive branch describes the exterior.  After we found these solutions, we realized that the positive branch was first derived by Abbassi \cite{Abbassi:1999wc}, and it was later criticized in Refs.~\cite{Faraoni:2017uzy, Faraoni:2021nhi}. On the latter criticism, we agree that, since these new coordinates are just equivalent to Kottler's coordinates apart from a diffeomorphism, there must be no physical differences; and hence no impact on AGN, as claimed in that work. On the other hand, Ref.~\cite{Faraoni:2017uzy} also claims that Abbassi's solution is not a true vacuum solution, since it requires (besides a cosmological constant) a non-null energy momentum tensor with a radial flux of matter. We disagree with this last statement, and we explicitly show here how to build this solution, how to do a coordinate transformation from Kottler's solution and we compute geometric quantities that characterize SdS spacetime; thus showing that such coordinates indeed describe a SdS spacetime. It is remarkable that it took so long for these SdS coordinates to be found and it is also remarkable that this solution is subtle to be verified using computer software; since it depends on a sequence of parameter conditions that, in part, are easier to be followed on paper. The codes for that verification and for generating the plots of this work are public \cite{rodrigues_SdS_Zenodo}.

Regarding the negative branch solution, it is not explicitly developed in \cite{Abbassi:1999wc} or commented in other references, but it received a quick comment in \cite{Abbassi:1999wc} dismissing it as unphysical; a comment which we naturally disagree with since it is just a different description, with the roles of time and space inverted, of the same SdS spacetime.

Besides finding the coordinate maps between all these mentioned SdS coordinate systems, we also uncover the map with respect to the maximally extended Lake-Israel coordinates of SdS spacetime \cite{Lake:2005bf}. Such map confirms (as expected) that neither the positive nor the negative branches are maximum extensions of SdS. This is expected since both are based on the Friedmann-Robertson-Walker (FRW) coordinates at large distances, and FRW coordinates on de Sitter spacetime describe a geodesically incomplete portion of the full de Sitter spacetime (e.g., \cite{Spradlin:2001pw}). For a time definition compatible with an expanding de Sitter regime (i.e., $\dot a > 0$), both branches describe a white hole (WH) and a standard cosmological horizon; while for a contracting de Sitter regime (i.e., $\dot a < 0$), they describe a BH together with an inverted cosmological horizon. As here shown, all the regions of the SdS Penrose diagram \cite{Gibbons:1977mu, Bousso:2002fq, Spradlin:2001pw, Lake:2005bf} can be explored in this coordinate system only by considering two sets of geodesics, one with a given cosmic time direction, and the other with its inverse.

For the geometrical conventions in this work, we follow Wald \cite{Wald:1984rg}. In particular we use the signature $\begin{pmatrix} - & + & + & + \end{pmatrix}$ and abstract index notation, with indices $a, b, c, d$ associated with spacetime.

\section{Schwarzschild-de Sitter spacetime: characterization} \label{sec:geometricCharacterization}
We start this section by briefly characterizing the de Sitter and Schwarzschild spacetimes. This is helpful to clarify our definitions. The de Sitter spacetime can be characterized by the following constant Riemann tensor \cite{StephaniExact},
\begin{equation} \label{eq:deSitterDef}
    R^{ab}_{\;\;\;\;cd} = \frac{\Lambda}{3}(\delta^a_c\delta^b_d - \delta^a_d\delta^b_c) \, ,  
\end{equation}
where $\Lambda$ is a positive constant (the cosmological constant). The above implies that: $R_{ab} = - \Lambda g_{ab}$ and the Kretschman scalar is $K = 8 \Lambda^2/3$. 

Due to the Birkhoff theorem, the Schwarzschild spacetime can be characterized as being a spherically symmetric solution of Einstein's vacuum field equation with $\Lambda =0$ (and the latter equation is equivalent to $R_{ab} = 0$). Apart from the trivial solution given by Minkowski spacetime, Scharzschild spacetimes have a Kretschman scalar that diverges in the center of the spherical symmetry.

Before characterizing the SdS, we recall the Kottler line element \cite{Kottler:1918cxc, StephaniExact, Faraoni:2017uzy} 
\begin{equation}\label{eq:Kottler}
  ds^2 = -f(r) \, dt^2 + \frac{1}{f(r)} \, dr^2 + r^2 d\Omega^2 \, ,
\end{equation}
with
\begin{equation} \label{eq:f}
    f(r) =  1 - \frac{2m}{r} - r^2 \frac{\Lambda}{3} \, ,
\end{equation}
where $m$ and $\Lambda$ are constants that are interpreted as the mass of the BH and the cosmological constant. 

In Kottler's coordinate system, there are two coordinate singularities corresponding to the two positive roots of $f(r)$. These are interpreted as the BH event horizon and the cosmological horizon. 
Freely falling observers may cross such horizons without detecting them by any local experiment \cite{Wald:1984rg}.  In both pure de Sitter and Schwarzschild spacetimes, one can introduce coordinate systems that eliminate the coordinate singularities. For example, for de Sitter, the cosmological (time-dependent) coordinates; and for Schwarzschild, the Kruskal–Szekeres \cite{Kruskal:1959vx, Szekeres:1960gm} and Israel \cite{Israel:1966zz} charts.  In the SdS case, Lake \cite{Lake:2005bf} built a regular, maximally extended coordinate system based on Israel’s approach.

The Birkhoff theorem for the Schwarzschild spacetime can be extended to consider a cosmological constant. In this case, paraphrasing Rindler \cite{1998PhLA..245..363R}: Every spherically symmetric solution of the vacuum field equations $R_{ab} = \Lambda g_{ab}$ is either equivalent to Kottler's generalization of the Schwarzschild space \eqref{eq:Kottler} or to the Bertotti-Kasner space. 

The Bertotti-Kasner spacetime describes a homogeneous universe with spherical symmetry. A simple geometrical way to differentiate between these spaces is provided by the Kretschmann scalar: For the Kottler line element, 
\begin{equation} \label{eq:KSdS}
     K = 48 \frac{m^2}{r^6} + 8 \frac{\Lambda^2}{3} \, ,
\end{equation} 
while for Bertotti-Kasner this scalar is a constant, $K = 8 \Lambda^2$. In conclusion, any spherically symmetric solution of $R_{ab} = \Lambda g_{ab}$ whose Kretschmann scalar is not constant will be diffeomorphic to the spacetime described by Eq.~\eqref{eq:Kottler}. Any spacetime diffeomorphic to the latter line element is here called a SdS spacetime.

\section{The regular FRW-based coordinates for SdS} \label{sec:building}
We consider a  spacetime with spherical symmetry and introduce coordinates such that the line element reads
\begin{equation} \label{eq:lineElement}
    ds^2 = a^2(\eta) \left[ -e^{\alpha(\eta, r)} \, d\eta^2 + e^{\beta(\eta, r)} \, dr^2 + r^2 \, d\Omega^2 \right].
\end{equation}
Here, \( a(\eta) \) is the cosmological scale factor, \( \alpha(\eta, r) \) and \( \beta(\eta, r) \) are  functions that depend on  the conformal time \( \eta \) and the radial coordinate \( r \). We note that  the above line element has been used recently to discuss the cosmological coupling of local objects \cite{Cadoni:2023lum, Cadoni:2024jxy}, but here we do not assume $\partial_\eta \alpha = 0$. On the other hand, we restrict the cosmological contribution to be pure de Sitter.

With the above line element, the Einstein field equations become
\begin{subequations} \label{eq:fieldEqs1}
\begin{align}
    & \frac{\dot{a}}{a} \alpha' + \frac{\dot{\beta}}{r} = 0, \label{eq:fieldEqs1a}\\[0.3cm]
    & 3 \frac{\dot{a}^2}{a^2} + \frac{e^{\alpha - \beta}}{r^2} \left(-1 + e^{\beta} + r\beta' \right) + \frac{\dot{a}}{a} \dot{\beta} =  e^{\alpha} a^2 \Lambda, \\[0.3cm]
    & \frac{\dot{a}^2}{a^2} e^{\beta - \alpha} + \frac{1 - e^{\beta} + r\alpha'}{r^2} + \\[0.3cm]
    & \hspace*{2cm} + e^{\beta - \alpha} \left(-2 \frac{\ddot{a}}{a} + \frac{\dot{a}}{a} \dot{\alpha} \right)= -e^{\beta} a^2 \Lambda \, , \nonumber
\label{pconserv}
\end{align}
\end{subequations}
where, for any quantity $X$, $\partial_\eta X = \dot X $ and $\partial_r X  =  X' $. If $\dot \alpha =0$, the first equation above can be promptly integrated in time, which is the route taken in Ref.~\cite{Cadoni:2024jxy}, for instance.

The cosmological scale factor is here fixed such that it satisfies the Friedmann equation, 
\begin{equation}\label{eq:aEqs}
    \frac{3 \dot{a}^2}{a^2} =  a^2 \Lambda \, .
\end{equation}

Substituting the above into \eqref{eq:fieldEqs1}, and eliminating $\dot \beta$ by using Eq.~\eqref{eq:fieldEqs1a},
\begin{subequations} \label{eq:fieldEqs3}
\begin{align}
  & \frac{\dot{a}}{a} \alpha' + \frac{\dot{\beta}}{r} = 0, \\
  &\frac{1 - e^{-\beta} + r \beta' e^{-\beta}}{r^2} =   a^2  \Lambda \left(1 - e^{-\alpha} + \frac{r\alpha'}{3} e^{-\alpha} \right)  , \\
  & \frac{e^{-\beta} + r e^{-\beta} \alpha' - 1}{r^2} +e^{-\alpha}\frac{\dot{a}}{a}\dot{\alpha}=   a^2  \Lambda \left( e^{-\alpha} -1\right) \, ,
\end{align}
\end{subequations}
or, equivalently,
\begin{subequations} \label{eq:fieldEqs4}
\begin{align} 
    & \frac{\dot{a}}{a} \alpha' + \frac{\dot{\beta}}{r} = 0, \label{eq:fieldEqs4a} \\[0.3cm]
    &  \frac{1}{  a^2 r^2} \partial_r \left[ r \left(1 - e^{-\beta}\right) \right] - \frac{\Lambda}{3 r^2} \partial_r \left[ r^3 \left(1 - e^{-\alpha}\right) \right]=0, \label{eq:fieldEqs4b} \\[0.3cm]
    &  \frac{1}{  a^2} \frac{1}{r^2} \left[ e^{-\beta} \left(1 + r\alpha' \right) - 1 \right] +\frac{\dot{a}}{  a^3}e^{-\alpha}\dot{\alpha}+\label{eq:fieldEqs4c} \\ 
    &\hspace*{4cm} +\Lambda \left(1 - e^{-\alpha} \right)=0.  \nonumber
\end{align}
\end{subequations}

Equation \eqref{eq:fieldEqs4b} can be integrated with respect to \( r \) to obtain
\begin{equation}\label{eq:7}
    2  m(\eta) = a r \left(1 - e^{-\beta}\right) - \frac{\Lambda}{3} a^3 r^3 \left(1 - e^{-\alpha}\right)\, ,
\end{equation}
where \( m(\eta) \) is the integration constant with respect to $r$. For $\Lambda = 0$, the solution is the Schwarzschild one, with $m$ and $ar$ as the Schwarzschild mass and radial coordinate, respectively.

From the above and integrating Eq.~\eqref{eq:fieldEqs4a} with respect to $\eta$, see Appendix \ref{app:expB},  $m(\eta)$ is found to be a constant, which is henceforth denoted by $m$, and 
\begin{equation} \label{eq:eBetaSol}
    e^{-\beta} = 1 - \frac{2m}{a r} + a^2 r^2 \frac{\Lambda}{3}   \left(e^{-\alpha} - 1\right).
\end{equation}

Since $r$ only appears explicitly inside the term $a r$, we define
\begin{equation}
    \ell \equiv a r \, ,
\end{equation}
and we look for solutions with $\alpha(r, \eta) = \alpha(\ell)$. From Eq.~\eqref{eq:eBetaSol}, one finds $\beta(r, \eta) = \beta(\ell)$. Hence, using Eq.~\eqref{eq:fieldEqs4a}, $\partial_{\ell} (\alpha + \beta) = 0$, which implies that  $\alpha + \beta$ is a constant. By a coordinate redefinition, one can remove this constant. Therefore, we find
\begin{equation}
    \alpha = - \beta \, .
\end{equation}

With the above, now one has to solve
\begin{equation}\label{eq:21}
    e^{\alpha} = 1 - \frac{2m}{\ell} + \ell^2 \frac{\Lambda}{3}  \left(e^{-\alpha} - 1\right) = f(\ell) + \ell^2 \frac{\Lambda}{3} e^{-\alpha}\, .
\end{equation}
The solution can be promptly found as
\begin{equation} \label{eq:ealphapm}
    e^{\alpha_{\pm}} \equiv A_\pm =\frac{1}{2} \left( f(\ell) \pm \sqrt{f^2(\ell) + 4 \ell^2 \frac{\Lambda}{3}   } \right) \, ,
\end{equation}
where $f$ is defined in Eq.~\eqref{eq:f}. It is important to note that \textit{neither $A_+$ nor $A_-$ have real roots}. Indeed, $A_{\pm}$ has the form $X \pm \sqrt{ X^2 + K}$, with $K > 0$, thus $A_- < 0$ and $A_+ > 0$ for any $\ell$.

With the above, we have found the complete solution, which reads
\begin{equation} \label{eq:ds2sol}
    ds_\pm^2 = a^2 \left( - A_\pm d\eta^2 + A^{-1}_{\pm} dr^2 + r^2 d\Omega^2\right) \, ,
\end{equation}
where $A_{\pm}$ can assume the solutions in Eq.~\eqref{eq:ealphapm}, $\ell = a(\eta) r$ and 
\begin{equation}
  a(\eta) = - \sqrt{\frac{3}{\Lambda}}\frac{1}{\eta} \, , 
\end{equation}
with $-\infty < \eta < 0$, which is a cosmological scale factor solution for de Sitter in conformal time \eqref{eq:aEqs}. 
It is also possible to use the physical time $t$ instead of the conformal time $\eta$ by introducing $dt^2 = a^2 d\eta^2$. 

In Ref.~\cite{Abbassi:1999wc}, only the case $A_+$, from Eq.~\eqref{eq:ealphapm} was considered to be acceptable.\footnote{See Eq.~(28) from that reference.} Indeed, $A_+$ is the easiest case to interpret. As we develop in the next sections, the negative branch solution, which depends on $A_-$, is  also a valid SdS description.

\begin{figure}
	\begin{tikzpicture}
      	\node (img1)  {\includegraphics[width=0.47\textwidth]{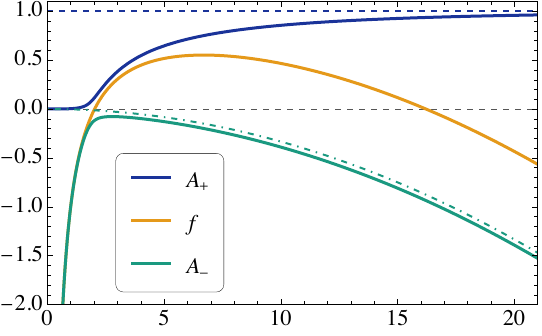}};
    	\node[below=of img1, node distance=0cm, yshift=1.1cm, xshift=0.3cm, font=\color{black}] {\large $\ell \, (m)$};
	\end{tikzpicture} 
	\caption{The plot compares the functions $A_+$ (solid blue), $A_-$ (solid green) \eqref{eq:ealphapm}  and $f$ (solid yellow) \eqref{eq:f}. They are written  as functions of the distance $\ell$, which is expressed in units of the BH mass $m$. Only $f$ has roots, $A_+$ is always positive and $A_-$ is always negative. The figure also shows the large $\ell$ asymptotics for  $A_+$ (blue dashed line, which is constant and equal to 1), and for $A_-$ (the green dot-dashed line, which decays proportionally with $ - \Lambda \ell^{2}$). This plot was done with $\Lambda  = 10^{-2} m^{-2}$. }
	\label{fig:plotAlphas} 
\end{figure}

The curves described by $A_\pm(\ell)$ and $f(\ell)$ can be plotted and numerically compared for fixed relation between $\Lambda$ and $m$. We focus on the most physically interesting case which is $m^2\Lambda \ll 1$. In Fig.~\ref{fig:plotAlphas} we plot and compare these curves.

\section{The geometric content and limiting cases} \label{sec:limitingCases}

\subsection{Geometric content} \label{sec:geometricContent}

One can directly verify from the line element \eqref{eq:ds2sol} that $R_a^b = \Lambda \delta_a^b$ for both $ds^2_+$ and $ds^2_-$. This confirms that the derived solution is indeed a vacuum solution with a cosmological constant. Moreover, the Kretschmann scalar can be straightforwardly computed, and it is found that it is the same of Eq.~\eqref{eq:KSdS}, but with $\ell$ in place of $r$. Therefore, considering the de Sitter-extended Birkhoff theorem \cite{1998PhLA..245..363R}, the line element \eqref{eq:ds2sol} with either $A_+$ or $A_-$ is a valid description of the SdS spacetime. For completeness, we state here the Riemann tensor for these coordinates,
\begin{equation}
    R^{\mu \nu}_{\;\; \lambda \kappa} = \frac{m}{\ell^3} A^{\mu \nu}_{\lambda \kappa}  + \frac{\Lambda}{3} (\delta^\mu_\lambda \delta^\nu_\kappa - \delta^\mu_\kappa\delta^\nu_\lambda) \, ,
\end{equation}
where $A^{\mu \nu}_{\lambda \kappa}$ is a rank-4 constant array that can only assume the following values: $\{0, \pm1, \pm2\}$. The result above is also valid for Kottler coordinates, with $\ell$ replaced by $r$. Further details on the computation of these geometric quantities are provided in code form \cite{rodrigues_SdS_Zenodo}.  The explicit coordinate transformation between the corresponding metrics is presented in the next section.

\subsection{$\Lambda = 0$ case}
We start by exploring the solution \eqref{eq:ds2sol} for $\Lambda = 0$. For this case, $a(\eta)$ is a constant that can be absorbed by coordinate redefinitions, namely $a d\eta \rightarrow d\eta$ and $a r \rightarrow r$. The expression for $A_\pm$ can be written as
\begin{equation}
    A_\pm = \frac{1}{2} \left ( 1 - \frac{2 m}{r} \pm \left | 1 - \frac{2 m}{r} \right | \right ) \, .
\end{equation}
For this $\Lambda = 0$ case, neither $A_+$ nor $A_-$ describes the Schwarzschild solution, since $A_+ = 0$ for $r< 2m$ and $A_- = 0$ for $r > 2m$. However,  by combining the non-null cases, one recovers the Schwarzschild solution. This since 
\begin{align}
      A_- & = 1 - \frac{2 m}{r} \mbox{ for } r < 2 m  \, ,\nonumber \\[.2cm]
      A_+ & = 1 - \frac{2 m}{r} \mbox{ for } r > 2 m \, . \label{eq:alphaSchwarzschild}
\end{align}

This  case shows that the proposed coordinates are not strictly valid for $\Lambda = 0$, since neither $ds_+$ nor $ds_-$ can individually reproduce Schwarzschild spacetime. In the next sections, for $\Lambda > 0$, it will be shown that both $A_+$ and $A_-$ lead to valid SdS coordinates.

\subsection{$m=0$ case} \label{sec:m=0}
For $m =0$, the solutions from Eq.~\eqref{eq:ealphapm} are
\begin{align}
    A_+ & = 1 \, , \nonumber \\
    A_- & = - \frac{\Lambda}{3} \ell^2 \, .
\end{align}
Both the $A_+$ and $A_-$ cases, once inserted into Eq.~\eqref{eq:ds2sol}, lead to de Sitter spacetime line elements. The $A_+$ branch is the easiest to understand, since it promptly expresses (a patch of) de Sitter spacetime in standard cosmological coordinates. This limiting situation is also obtained asymptotically when $\ell \to \infty$, i.e., when considering distances much larger than the local scale of the compact object.

The line element of the $A_-$ solution reads
\begin{equation}
    ds^2 = - \frac{3}{\Lambda r^2} dr^2 + a^4 r^2 \frac{\Lambda}{3} d\eta^2 + a^2 r^2 d\Omega^2 \, ,
\end{equation}
where we changed the order of the $dr$ and $d\eta$ terms, since, in this case, $r$ is the time coordinate. A straightforward way to verify the geometric nature of the corresponding spacetime is to compute the Riemann tensor. From the latter one verifies that it satisfies Eq.~\eqref{eq:deSitterDef}, implying that it indeed describes a de Sitter spacetime. Another useful verification procedure is to introduce the following coordinates,
\begin{equation} \label{eq:map+-}
    \tilde r = - \frac{1}{\Lambda \eta} \; \mbox{ and } \; \tilde \eta = - \frac{1}{\Lambda r}\, .
\end{equation}
This implies that 
\begin{equation}
    \ell = a(\eta) r = - \sqrt{\frac{3}{\Lambda}} \frac{r}{\eta} = - \sqrt{\frac{3}{\Lambda}} \frac{\tilde r}{\tilde \eta} = a(\tilde \eta) \tilde r\, ,
\end{equation}
that is, $\ell$ preserves its form under this coordinate transformation. Moreover,
\begin{align}
    & \frac{3}{\Lambda} \frac{dr^2}{r^2}  = \frac{3}{\Lambda} \frac{d\tilde \eta^2}{\tilde \eta^2} = a^2(\tilde \eta) d\tilde \eta^2 \, , \\[.5cm] 
    & \frac{3}{\Lambda}\frac{r^2}{ \eta^2} \frac{d \eta^2}{\eta^2} = \frac{3}{\Lambda} \frac{\tilde r^2}{\tilde \eta^2} \frac{d\tilde r^2}{\tilde r^2} = a^2(\tilde \eta) d\tilde r^2 \, .
\end{align}
In conclusion, the transformation \eqref{eq:map+-} maps the de Sitter line element obtained from the $A_-$ solution to that of the $A_+$ solution.

\subsection{Newtonian limit}
In order to improve physical intuition in a physical setting in which neither $\Lambda$  or $m$ are zero, we evaluate how to achieve the Newtonian limit. The main point here is understanding the negative branch solution.

The Newtonian limit of the $ds^2_+$ case is straightforward. First we redefine time from the conformal to the physical one ($dt = a(\eta) d\eta$), then, expanding in powers of $m$ and neglecting $\Lambda$ terms,
\begin{equation}
    g_{00}^{(+)} = - A_+ \approx -  \left(1 - \frac{2 m}{\ell} \right) \, .
\end{equation}
The Newtonian potential (with $G = c = 1$) is identified as $-m/\ell$. Physically, in this limit, the gradient of the Newtonian potential gives the acceleration of a test particle.

For the line element $ds^2_-$, the time-time component is not $g^{(-)}_{00}$, but $g^{(-)}_{11}$. 
The radius related to the spherical symmetry is still given by $\ell = r a(\eta)$, but $r$ is the time coordinate and $\eta$ is space-like. Thus, the time-time metric component reads,
\begin{align}
    g_{11}^{(-)} & =  - \frac{a^2}{2} \left( \sqrt{f^2(\ell) + 4 \ell^2 \frac{\Lambda}{3}}  - f(\ell)\right)^{-1}\nonumber \\[.1cm]
    & = - \frac{a^2}{2} \frac{3}{4 \ell^2 \Lambda} \left( \sqrt{f^2(\ell) + 4 \ell^2 \frac{\Lambda}{3}}  + f(\ell) \right) \nonumber  \\[.1cm]
    & \approx  - \frac{3}{4 r^2 \Lambda} \left(1 - \frac{2 m}{\ell}\right) \, .
\end{align}
By redefining the time\footnote{Namely, $\frac{\sqrt 3}{2 r \sqrt \Lambda}dr = d\tau$, from which one finds $r = r(\tau)$ and $g_{11}^{(-)} = - \left(1 - \frac{2 m}{\ell}\right)$.} $r$, we identify the Newtonian potential, again, as $-m/\ell$. Hence, both the positive and the negative branch solutions have the same Newtonian limit (apart from coordinate redefinitions).

\section{Direct coordinate transformations}

\subsection{Kottler coordinates} \label{sec:kottler}

In this section, we explicitly present the transformation map between the line elements \eqref{eq:Kottler} and \eqref{eq:ds2sol}. The process of deriving this map between coordinates provides an independent demonstration that the regular, FRW-based line element discussed here indeed describes the SdS spacetime. It is possible to find such a coordinate transformation for both the positive and the negative branches simultaneously, since the steps are the same in both cases. We start from the line element $ds^2_{\pm}$ \eqref{eq:ds2sol}.

Using the physical time ($dt  = a \, d\eta $) and $ r = \ell / a$,  Eq.~\eqref{eq:ds2sol} becomes
\begin{equation}
\begin{split}
    ds^2_\pm = & -\left(A_\pm - \frac{\Lambda}{3} \frac{\ell^2}{A_\pm} \right) \, dt^2 
    + \frac{1}{A_{\pm}} \, d\ell^2 \\[.1cm]
    & - 2\sqrt{\frac{\Lambda}{3}} \ell \frac{1}{A_\pm} \, d\ell \, dt 
    + \ell^2 \, d\Omega^2 \, .
\end{split}
\end{equation}
In the above, it was also used that $\partial_t a/a = \sqrt{\Lambda/3}$.  From Eq.~\eqref{eq:21}, 
\begin{equation}\label{eq:29}
    ds^2_\pm = -f \, dt^2 + \frac{d\ell^2}{A_\pm} - 2\sqrt{\frac{\Lambda}{3} } \frac{\ell}{A_\pm} \, d\ell \, dt + \ell^2 \, d\Omega^2 \, .
\end{equation}

The last step consists of the elimination of the $d\ell dt$ term. To this end, we rewrite the line element as
\begin{align}
    ds^2_\pm & = -f \left(dt + \sqrt{\frac{\Lambda}{3}} \frac{\ell}{f A_\pm}  \, d\ell\right)^2 + \nonumber \\
    & + \left(1 + \frac{\Lambda}{3}  \frac{\ell^2}{f A_\pm}\right) \, \frac{d\ell^2}{A_\pm} + \ell^2 \, d\Omega^2 \, .
\end{align}
Using Eq.~\eqref{eq:21} to simplify the $d\ell^2$ term, 
\begin{equation}
    ds^2 = -f(\ell) \, d\tilde t^2 + \frac{1}{f(\ell)} \, d\ell^2 + \ell^2 \, d\Omega^2 \, .
\end{equation}
The above line element is the the Kottler one \eqref{eq:Kottler}, thus we have achieved the desired coordinate transformation.

In conclusion, the explicit mapping of the coordinates between the line elements \eqref{eq:Kottler} and \eqref{eq:ds2sol} reads
\begin{equation}\label{eq:CoordinateTransformations}
    \ell = a \,  r \qquad \mbox{ and } \qquad \pm d\tilde{t} = a \, d\eta + \sqrt{\frac{\Lambda}{3}}   \,  \frac{\ell}{f A_\pm} \, d\ell \, .
\end{equation}
The expression for $d\tilde{t}$ defines a locally exact differential. In the above, we stress that the map with $A_+$ should be used for $ds^2_+$, while the map with $A_-$ should be used for $ds^2_-$. The sign in front of $d\tilde t$ is unrelated with the sign of $A_\pm$. Had we considered a contracting universe, the sign in from of $\sqrt{\Lambda/3}$ would be inverted. As expected, the map with Kottler coordinates is not valid at $f = 0$.

The case with $A_+$ was first derived by Abbassi\footnote{This coordinate transformation, as presented in  \cite{Abbassi:1999wc} and here re-derived and extended,  was contested in \cite{Faraoni:2017uzy}. The discrepancy seems related to a factor 2 in the nondiagonal term in their Eq.~(3.8), which together with their eqs.~(3.6) and (3.9), can be compared with our Eq.~\eqref{eq:CoordinateTransformations}. One can directly verify that $R_{ab} = \Lambda g_{ab}$ for Eq.~\eqref{eq:ds2sol}, but this is not the case of the line element in Eq.~(3.10) from \cite{Faraoni:2017uzy}, which uses a different  transformation.} \cite{Abbassi:1999wc}. Here we are showing in detail such derivation and extending it to the negative branch solution.

\subsection{The map between the positive and negative branch solutions} \label{sec:mapPositiveNegative}

The negative branch solution reads \eqref{eq:ds2sol}
\begin{equation} \label{eq:ds2solminus}
    ds_-^2 =  - a^2 |A^{-1}_-| dr^2 + a^2 |A_-| d\eta^2 +  \ell^2 d\Omega^2\, ,
\end{equation}
with $a = a(\eta)$, $A_- = A_-(\ell)$ and $\ell = a r$.

From the definition \eqref{eq:ealphapm}, the function $A_-$ can be related to $A_+$ as follows,
\begin{equation}
    A_-^{-1} = - \frac{1}{\ell^2} \frac{3}{\Lambda} A_+ \, .
\end{equation}
Hence,
\begin{equation}
    ds^2_-  = -\frac{3}{\Lambda r^2} A_+ dr^2 + a^2 \ell^2 \frac{\Lambda}{3} A_+^{-1} d\eta^2 + \ell^2 d\Omega^2 \, .
\end{equation}

Clearly $r$ and $\eta$ in the above line element have the geometric nature of time and space respectively. This suggests using a map similar to the one used in the pure de Sitter case, Eq.~\eqref{eq:map+-}. The latter map, although derived for a particular case, is already sufficient for the full SdS case, namely
\begin{align}
    ds^2_- & = - a^2(\tilde \eta) A_+ d\tilde \eta^2 + a^2(\tilde \eta) A_+^{-1} d\tilde r^2 + \ell^2 d\Omega^2 \nonumber \\[.3cm] 
    & = ds^2_+\, ,
\end{align}
Hence, the map between the negative and positive branch solutions is the same of Eq.~\eqref{eq:map+-},
\begin{equation}
    \tilde r = - \frac{1}{\Lambda \eta} \; \mbox{ and } \; \tilde \eta = - \frac{1}{\Lambda r}\, . \tag{\ref{eq:map+-}} 
\end{equation}
We stress that this map depends explicitly on $\Lambda$, implying that it is not valid in the Schwarzschild case, thus in agreement with the results from Sec.~\ref{sec:m=0}.

Henceforth, we will only use $A_+$ and the line element \( ds^2_+ \), which are now denoted without the positive sign, thus,
\begin{equation} \label{eq:ds^2} 
    ds^2  = a^2(\eta) \left( - A d\eta^2+ A^{-1} dr^2 + r^2 d\Omega^2\right)  \, .
\end{equation}

\subsection{Maximally extended Lake-Israel coordinates} \label{sec:lake}

More recently, Lake~\cite{Lake:2005bf} extended Israel coordinates to the SdS spacetime. In these maximally extended coordinates, the horizons are regular and there are ingoing and outgoing geodesics that can cross them. The corresponding line element is given by
\begin{equation} \label{eq:ds2Lake}
    ds^2 = g(u, w)\, du^2 + 2\, du\, dw + v^2(u, w)\, d\Omega^2 \,,
\end{equation}
where
\begin{align}
    g(u, w) &= \frac{w}{3 u v C^2} \Big[ -2C(C - v)^2 + u w (2C + v) \Big], \\[.2cm]
    v(u, w) &= \frac{u w (3m - C)}{C^2} + C \, .
\end{align}

The cosmological constant \( \Lambda \) is related to the parameters \( C \) and \( m \) by
\begin{equation}
    \Lambda = \frac{3(C - 2m)}{C^3} \, .
\end{equation}
In the SdS case, $C$ satisfies \( 2m < C < 3m \), with \( m \) interpreted as the central mass.

Radial null geodesics can be parameterized with $w$ and they satisfy either $du=0$ or $du/dw = - 2/g$ \cite{Lake:2005bf}.

One disadvantage of this coordinate system is that it is hard to understand intuitively. The correspondence of $w$ with the cosmological time $t$ or $\eta$ is far from straightforward. We explore this connection in what follows.

The local correspondence between \eqref{eq:ds2Lake} and the SdS cosmological coordinates \eqref{eq:ds^2} is derived in Appendix \ref{app:lake}, whose result reads
\begin{subequations} \label{eq:coord_map}
\begin{align}
    r a(t) & = v(u,w) \, , \\[.2cm]
    dt & =  \left[ \pm\left(1 - \frac{f(v)}{\partial_u v \partial_wv} \right) - \sqrt{\frac{\Lambda}{3}} \frac{v}{A(v)} \right] \frac{\partial_u v}{f(v)}du \nonumber \\
    & +\left( \pm 1 - \sqrt{\frac{\Lambda}{3}} \frac{v}{A(v)} \right) \frac{\partial_w v}{f(v)} dw \, .
\end{align}
\end{subequations}
For a universe that is contracting, $\dot a < 0$, the signs in front of the $\sqrt{\Lambda/3}$ term should be inverted. 

While both coordinate systems are regular at the horizons, the coordinate map relating them is not, as shown in Appendix \ref{app:lake}. This indicates, that although the line elements \eqref{eq:ds2Lake} and \eqref{eq:ds^2} can be locally related by coordinate transformations, the map is not globally valid. In particular, we find that the sign choices in Eq.~\eqref{eq:coord_map} must vary between different geodesics. Additional details are given in Appendix \ref{app:lake}.

In the next section, we explore the causal structure in SdS from the perspective of the SdS cosmological coordinates here explored.

\section{Causal Structure Analysis} \label{sec:causal}

\begin{center}
    \begin{figure}
    \begin{tikzpicture}[scale=4]
    
    \coordinate (O) at (0,0);
    \coordinate (C) at (1,1);
    \coordinate (D) at (2,0);
    \coordinate (E) at (1,-1);
    \coordinate (OC) at (0,1);
    \coordinate (OE) at (0,-1);
    \coordinate (CD) at (2,1);
    \coordinate (DE) at (2,-1);
    
    \draw[gray, thick] (O) -- (C) -- (D) -- (E) -- cycle;
    
    \draw[thick,decorate,decoration={snake, segment length=4pt, amplitude=1.5pt}] (OC) -- (C);
    \draw[thick,decorate,decoration={snake, segment length=4pt, amplitude=1.5pt}] (OE) -- (E);
    
    \draw[thick] (C) -- (CD);
    \draw[thick] (DE) -- (E);
    
    \draw[dashed, gray] (DE) -- (CD);
    \draw[dashed, gray] (OE) -- (OC);
    
    \node[above] at (0.5,1) {\small $\ell=0$};
    \node[below] at (0.5,-1) {\small $\ell=0$};
    \node[above] at (1.5,1) {\small $\ell=\infty$};
    \node[below] at (1.5,-1) {\small $\ell=\infty$};
    
    \node at (1,0) {\small I};
    \node at (0.25,-0.6) {\small II};
    \node at (0.25,0.6) {\small III};
    \node at (1.75,0.6) {\small IV};
    \node at (1.75,-0.6) {\small V};
    
    \draw[RoyalBlue, line width=1.7pt, -stealth] (0, -0.5) -- (1.5, 1);
    \draw[RoyalBlue, line width=1.7pt, -stealth] (0, -0.5) -- (0.75, 0.25);
    
    \draw[RoyalBlue, line width=1.7pt, -stealth] (0.5, -1) -- (2, 0.5);
    \draw[RoyalBlue, line width=1.7pt, -stealth] (0.5, -1) -- (1.25, -0.25);
    
    \draw[ForestGreen, line width=1.7pt, -stealth] (1.5, -1) -- (0, 0.5);
    \draw[ForestGreen, line width=1.7pt, -stealth] (1.5, -1) -- (0.75, -0.25);
    
    \draw[ForestGreen, line width=1.7pt, -stealth] (2, -0.5) -- (0.5, 1);
    \draw[ForestGreen, line width=1.7pt, -stealth] (2, -0.5) -- (1.25, 0.25);
    
    \end{tikzpicture}
    \caption{Penrose diagram for SdS spacetime centered at the ``standard universe'' (region I). The horizons are depicted by solid gray lines (without arrows). The diagram can be continued to the left or to the right \cite{Gibbons:1977mu, Bousso:2002fq, Lake:2005bf}. Region II is a white hole (WH) interior, region III is a BH interior, region IV is a future region beyond the cosmological horizon and region V is a past region beyond the cosmological horizon. The blue and green solid lines with arrows show the radial propagation of light: either away (blue) or toward (green) the singularity at $\ell=0$.} \label{fig:penroseDiagram}
    \end{figure}
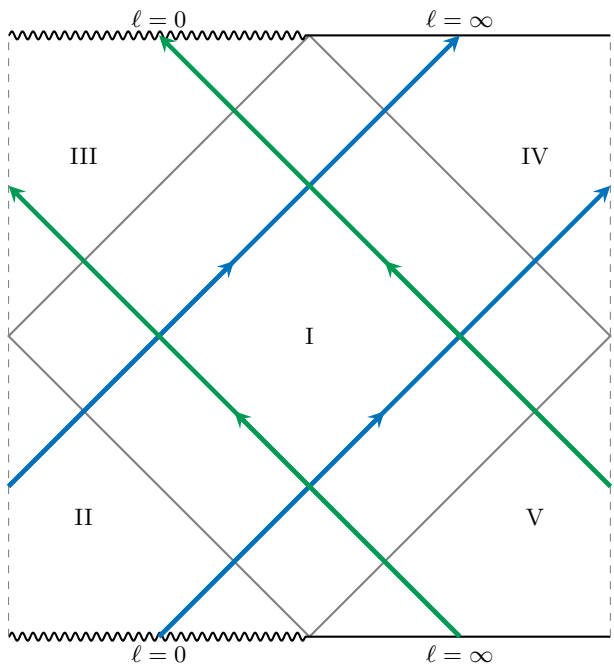
\end{center}
The causal structure of the SdS spacetime has been investigated in several works~\cite{Gibbons:1977mu, Lake:1977ui, 1986NCimB..91..126B, Lake:2005bf, Uzan:2010nw}. The global causal structure can be understood through the Penrose diagram shown in Fig.~\ref{fig:penroseDiagram}, which represents the maximal extension of the Kottler metric. In what follows, we analyze which regions of this maximal extension can be covered by the line element \eqref{eq:ds^2}.

\begin{figure}
	\begin{tikzpicture}
      	\node (img1)  {\includegraphics[width=0.45\textwidth]{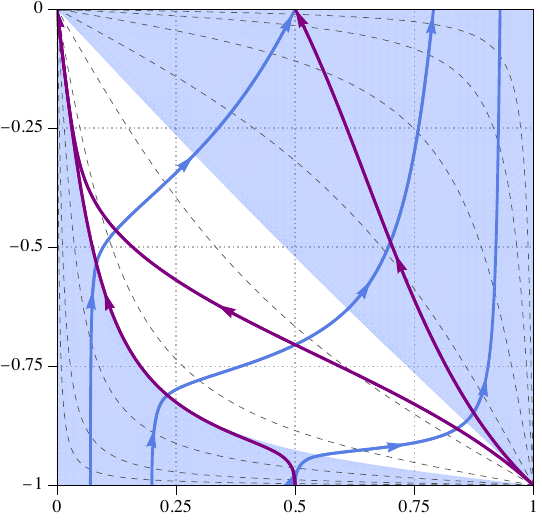}};
            \node[below=of img1, node distance=0cm, yshift=1.1cm, xshift=0.4cm, font=\color{black}] {\large $\bar r$};
            \node[left=of img1, node distance=0cm, xshift=0.8cm, yshift=0.5cm, rotate=90,  font=\color{black}] {\large $\bar \eta$};
	\end{tikzpicture} 
	\caption{Null geodesics of SdS spacetime using compactified versions of the $r$ and $\eta$ coordinates from \eqref{eq:ds^2}. The coordinates compactification is done with $\bar x = 2 \arctan(x)/\pi$. Ingoing (purple curves with arrows) and outgoing (blue curves with arrows) geodesics are defined from Eq.~\eqref{eq:detadlambda}, with different boundary conditions at $\bar r = 0.5$. Dashed gray lines represent curves of constant $\ell$. The physical singularity at $\ell = 0$ lies along the axes $\bar r =0$ and $\bar \eta = -1$. The two bluish regions correspond to values where $f(\ell) < 0$; their boundaries with the central white region mark the horizons, which are coordinate singularities in Kottler coordinates. The upper (bluish) boundary corresponds to the cosmological horizon, while the lower (bluish) boundary corresponds to the white-hole horizon. This plot was done using $m^2 \Lambda = 10^{-2}$ and $\dot a > 0$. } 
    \label{fig:plotEtaRgeodesics} 
\end{figure}

Let $\ell_1$ and $\ell_2$ be the two positive roots of $f(\ell)$, with $\ell_1 < \ell_2$. From $ds^2$ \eqref{eq:ds^2}, radial (i.e., $d\Omega = 0$) light-like geodesics parametrized with the affine
parameter $\lambda$ satisfy
\begin{equation} \label{eq:dt2dlambda2}
    \frac{d\eta^2}{d\lambda^2} =  \frac{dr^2}{d\lambda^2} A^{-2}\, .
\end{equation}

Given that $A > 0$, and considering $\lambda$ such that future-directed light-like geodesics satisfy $\frac{d\eta}{d\lambda} > 0$, this implies
\begin{equation} \label{eq:detadlambda}
    \frac{d\eta}{d\lambda} = \frac{1}{A} \left| \frac{dr}{d\lambda} \right| \, . 
\end{equation}
With respect to the radial coordinate $r$, ingoing and outgoing light rays, respectively, are defined from conditions $dr/d\lambda  <0$ and $dr/d\lambda  > 0$. In Fig.~\ref{fig:plotEtaRgeodesics} we use a compactified version of these coordinates, together with the above equation, to display the light-like geodesics in SdS space-time. 

At either the horizons $f(\ell) = 0$ and Eq.~\eqref{eq:detadlambda} implies that
\begin{equation} \label{eq:detadrHorizon1}
    \left. \frac{d\eta}{dr}\right|_{f=0} = \left. \pm \frac{1}{A}\right|_{f=0}  = \pm \frac{1}{\ell}\sqrt{\frac{3}{\Lambda}} \, ,
\end{equation}
where the $+$ and $-$ signs are respectively for the $r$-outgoing and $r$-ingoing light-rays. On the other hand, the horizons are at fixed values of $\ell$, which implies that, along any of the horizons, $r$ and $\eta$ satisfy $\partial_r( a(\eta) r) =0 $, thus
\begin{equation} \label{eq:detadrHorizon2}
    \left. \frac{d\eta}{dr} \right|_{f=0}  = - \frac{a}{\dot a r} = - \frac{1}{\ell} \sqrt{\frac{3}{\Lambda}}\, ,
\end{equation}
Where it was used that $\dot a > 0$. The sign of the above equation would be the opposite for $\dot a < 0$. We consider the first case.

The two previous equations define curves in the $r - \eta$ plane. Since they arise from first-order differential equations, if the curves intersect at a point where their derivatives coincide, they must be identical at all points. Equations \eqref{eq:detadrHorizon1} (for the $r$-ingoing light ray) and \eqref{eq:detadrHorizon2} are identical at the horizon. Consequently, the $r$-ingoing light ray can asymptotically approach the horizon but it cannot ever reach it. This behavior applies to both horizons, but only for the ingoing mode. It follows that the coordinates in the line element \eqref{eq:ds^2} are not maximally extended and cannot cover all regions of the Penrose diagram \ref{fig:penroseDiagram}.

In Fig.~\ref{fig:plotEtaRgeodesics} we plot numerical realizations of the null-geodesics in the same coordinates of the line element \eqref{eq:ds^2}, apart from a compactification. The plot assumes $\dot a > 0$ and shows that only the outgoing light rays can cross the horizons. This result is also valid for the ingoing light rays with respect to the coordinate $\ell$, as detailed below.

One can also define ingoing and outgoing geodesics with respect to the $\ell$ coordinate. This approach confirms that, using the coordinates \eqref{eq:ds^2}, the horizons can be crossed in a single direction; and also shows that only in the region $f(\ell) > 0$ there exist ingoing and outgoing modes: the other regions are only compatible with increasing $\ell$. This is detailed in Appendix \ref{app:HC}. From this, we conclude that these coordinates cover regions I, II, and IV of Fig.~\ref{fig:penroseDiagram}. Considering the ingoing modes with respect to $\ell$, there is a single geodesics in Fig.~\ref{fig:plotEtaRgeodesics} with this property, the purple one between the horizons (region I). As shown in Appendix \ref{app:HC}, ingoing $\ell$-modes are only possible in the region between the horizons.

It is not surprising that we were unable to find a all the regions of SdS spacetime (Fig.~\ref{fig:penroseDiagram}) from the line element \eqref{eq:ds^2} with $\dot a >0$. Solutions with the restriction $\dot a > 0$ can only cover part of de Sitter spacetime. Considering the time orientation as either $\dot a >0$ or $\dot a < 0$ changes the behavior of the horizons. Indeed, by considering \( a(-\eta) \) in place of $a(\eta)$ in \eqref{eq:ds^2}, one describes a contracting universe in which light geodesics cross the horizons through the opposite orientation. Hence, by combining both coordinate sets, it is possible to cover all the regions I to V in Fig.~\ref{fig:penroseDiagram}. 

\section{Conclusions} \label{sec:conclusions}
Here we have carefully derived two regular coordinate systems describing the Schwarzschild-de Sitter (SdS) spacetime, based on Friedmann-Robertson-Walker (FRW) coordinates, as shown in Eq.~\eqref{eq:ds2sol}. In these coordinates, the only remaining singularity is the central one; both SdS horizons are regular. While other regular coordinate systems exist for SdS, the ones presented here are distinguished by their explicit dependence on cosmological time ($t$ or $\eta$), thereby directly linking the local time near the singularity to the cosmological evolution. Although physically intuitive and closely related to various proposals addressing the interplay between BHs and cosmology~\cite{Faraoni:2007es, Croker:2021duf, Farrah:2023opk, Cadoni:2023lqe, Amendola:2023ays, Cadoni:2023lum, Cadoni:2024jxy, Faraoni:2024ghi, Croker:2024jfg, Poplawski:2024rys, Cadoni:2024rri}, these coordinates lead to some unexpected consequences: only outgoing null geodesics can cross the horizons in finite cosmological time, while ingoing null geodesics cannot cross either horizons. This behavior is explicitly illustrated in Fig.~\ref{fig:plotEtaRgeodesics}. Compared to the approach in Ref.~\cite{Cadoni:2024jxy} and related works, the solutions discussed here require $\dot{\alpha} \ne 0$ in our starting line element~\eqref{eq:lineElement}, thus suggesting a new approach to the subject.

It is important to stress that the positive-branch solution in Eq.~\eqref{eq:ds2sol} was first found in Ref.~\cite{Abbassi:1999wc}, though it was accompanied by some incorrect conclusions regarding the associated physical phenomena and the dismissal of the negative-branch solution. These claims, with which we disagree, were later criticized in Refs.~\cite{Faraoni:2017uzy, Faraoni:2021nhi}. Interestingly, the latter works also entirely dismissed the solution of Ref.~\cite{Abbassi:1999wc} as incompatible with the SdS geometry, a conclusion we do not support.

In addition to a careful and independent re-derivation of these solutions from the Einstein equations (Sec.~\ref{sec:building}), we aim to have fully clarified the geometrical structure of the positive-branch solution (Secs.~\ref{sec:geometricCharacterization} and \ref{sec:geometricContent}), its mapping to Kottler coordinates (Sec.~\ref{sec:kottler}), the explicit equivalence between the positive and negative branches (Sec.~\ref{sec:mapPositiveNegative}), together with an explicit mapping to the maximally extended Lake-Israel coordinates (Sec.~\ref{sec:lake} and Appendix~\ref{app:lake}). We also analyze the causal structure directly in the SdS cosmological coordinates~\eqref{eq:ds^2} (Sec.~\ref{sec:causal}), leading to Fig.~\ref{fig:plotEtaRgeodesics} and its comparison with the Penrose diagram (Fig.~\ref{fig:penroseDiagram}) and the null geodesics in Lake-Israel coordinates (Appendix~\ref{app:lake}).

We expect that these conclusions regarding the SdS spacetime in different coordinate systems will be particularly useful for future developments in the study of BHs in different backgrounds, beyond the de Sitter case.

\begin{acknowledgments}

    We thank Valerio Faraoni and Mariano Cadoni for helpful discussions.
    LL thanks \textit{Fundação de Amparo à Pesquisa e Inovação do Espírito Santo} (FAPES, Brazil) for support. DCR thanks \textit{Centro Brasileiro de Pesquisas Físicas} (CBPF) and \textit{Núcleo de Informação C\&T e Biblioteca} (NIB/CBPF) for hospitality, where part of this work was done. He also acknowledges \textit{Conselho Nacional de Desenvolvimento Científico e Tecnológico} (CNPq, Brazil), FAPES (Brazil) and \textit{Fundação de Apoio ao Desenvolvimento da Computação Científica} (FACC, Brazil) for partial support. 
\end{acknowledgments}

\appendix

\section{On the viability of $\dot \alpha \not= 0$} \label{app:dotAlpha}

In Ref.~\cite{Cadoni:2020jxe} it is argued that the line element \eqref{eq:lineElement} together with an spherically symmetric energy-momentum tensor with $T_0^1=0$ would lead to the condition 
\begin{equation} \label{eq:dadb}
    \dot \alpha \dot \beta =0 \, .
\end{equation}
It is also shown that the case $\dot \beta = 0$ does not lead to local structures that are dynamically coupled to the cosmological background. Hence, solutions with dynamical complying would require $\dot \alpha =0$. Afterwards, some works assumed this condition (e.g., \cite{Cadoni:2023lqe, Cadoni:2024jxy}). 

We understand the condition $\dot \alpha = 0$ as a working hypothesis that lead to relevant results in the context of the coupling between compact objects and cosmology.  Here we review the demonstration of Ref.~\cite{Cadoni:2020jxe} and show that that Eq.~\eqref{eq:dadb} is is not necessary.

First we briefly note that the solution we present here \eqref{eq:ds^2} is already a counterexample, since for this solution $\alpha = - \beta$, $T_0^1 \propto \Lambda \delta_0^1 = 0$ and $\dot \alpha \not=0$. 

In Ref.~\cite{Cadoni:2020jxe}, the condition \eqref{eq:dadb} is obtained by using: $i$) the Einstein field equations for the line element \eqref{eq:lineElement}; $ii$) energy-momentum tensor conservation; $iii$) the condition of no radial flux of energy ($T^0_i=0$). The last condition imposes a restriction on the metric solutions [given by their Eq. (7)], namely
\begin{equation} \label{eq:t0irest}
    \frac{\dot a}{a} \alpha'+ \frac{\dot \beta }{r} = 0 \, .
\end{equation}
By using an arbitrary energy-momentum tensor compatible with the Einstein equations and the condition $T^0_i = 0$, neither conditions $i$ nor $ii$ can impose any constraint on the metric components. This since the Einstein equations, for arbitrary $T_{ab}$, do not restrict the metric and the energy-momentum conservation is an identity for the Einstein equations. Hence there is no further metric restriction to be found beyond Eq.~\eqref{eq:t0irest}.

In conclusion, there are no problems in considering cosmologically coupled solutions with the condition $\dot \alpha =0$, but this condition is not necessary.

\section{Derivation of $e^{-\beta}$} \label{app:expB}

Since, from Eq.~\eqref{eq:fieldEqs4a}, $\partial_\eta \left(e^{-\beta} a\right) = \dot{a}  e^{-\beta}(1+r\alpha')$,  Eq.~\eqref{eq:fieldEqs4c} can be written as
\begin{equation}
 \frac{1}{a^2} \frac{1}{r^2} \left[ \frac{1}{\dot{a}} \partial_\eta \left(e^{-\beta} a\right) - 1 \right]-\frac{\dot{a}}{ a^3}\partial_\eta e^{-\alpha} +\Lambda\left(1 - e^{-\alpha} \right)=0 \, ,
\end{equation}
or, equivalently,
\begin{equation}
\partial_\eta \left(e^{-\beta} a\right) = \dot{a} + r^2 \frac{\dot{a}^2}{a} \partial_\eta e^{-\alpha} -\Lambda r^2 \dot{a} a^2  \left(1 - e^{-\alpha}\right) \, .
\end{equation}

Integrating the previous equation, using Eq.~\eqref{eq:aEqs}, and introducing $h(r)$ as the integration constant with respect to $\eta$,
\begin{widetext}
\begin{align}
    e^{-\beta} &= 1 + \frac{r^2}{a} \int \frac{\dot{a}^2}{a} \partial_\eta e^{-\alpha} d\eta - \Lambda \frac{ r^2}{a} \int \dot{a} a^2 \left(1 - e^{-\alpha}\right) d\eta + \frac{h(r)}{a}     \nonumber  \\[.5cm]
    & = 1 + \frac{r^2}{a} \left( \frac{\dot{a}^2}{a} e^{-\alpha} - \int \partial_\eta \left(\frac{\dot{a}^2}{a} \right) e^{-\alpha} d\eta \right) - \Lambda \frac{ r^2}{a} \int \dot{a} a^2 \left(1 - e^{-\alpha}\right) d\eta + \frac{h(r)}{a}  \nonumber \\[.5cm]
    & = 1 + \frac{r^2}{a} \left(  \frac{\dot{a}^2}{a} e^{-\alpha} -\Lambda \int \dot{a} a^2 e^{-\alpha} d\eta   \right) - \Lambda \frac{ r^2}{a} \int \dot{a} a^2 \left(1 - e^{-\alpha}\right) d\eta + \frac{h(r)}{a}  \nonumber \\[.5cm]
    & = 1 + r^2 \frac{\dot{a}^2}{a^2} e^{-\alpha} -\Lambda \frac{r^2}{a}\int \dot{a} a^2 d\eta+ \frac{h(r)}{a} \nonumber \\[.5cm]
    & = 1 + a^2 r^2 \frac{\Lambda}{3}  \left(e^{-\alpha}-1\right) + \frac{h(r)}{a} \, . \label{eq:14}
\end{align}
\end{widetext}

Substituting the above into Eq.~\eqref{eq:7}, we find that $2 m(\eta) = - r h(r)$, hence $m$ and $r h$ are constants. And, in conclusion,
\begin{equation}
    e^{-\beta} = 1 - \frac{2m}{a r} + a^2 r^2 \frac{\Lambda}{3}   \left(e^{-\alpha} - 1\right) \, ,
\end{equation}
which is the answer in Eq.~\eqref{eq:eBetaSol}.

\section{Lake-Israel coordinates: map derivation and geodesics analysis} \label{app:lake}

To determine the transformation map between the line elements \eqref{eq:ds2Lake} and \eqref{eq:ds^2},  we use Kottler coordinates $(\tilde t, \ell)$ as an intermediate step. The relation between $\ell$ and $(u,w)$ is promptly found as $\ell = v(u,w)$. Assuming that $\tilde t$ can be written as functions of $u$ and $w$, and using $ f = f(\ell) = f(v(u,w))$, we find
\begin{align}  
ds^2 =& -\left[ f \left(\partial_u \tilde{t} \right)^2 
         - \frac{1}{f} \left(\partial_u v \right)^2 \right] du^2 \nonumber \\[.2cm]
       &+ \left[ -f \left(\partial_w \tilde{t} \right)^2 
         + \frac{1}{f} \left(\partial_w v \right)^2 \right] dw^2 \label{eq:ds2LakeKottler} \\[.2cm]
       &+ 2\left[ -f \left(\partial_u \tilde{t} \right)\left(\partial_w \tilde{t} \right) 
         + \frac{1}{f} \left(\partial_u v \right)\left(\partial_w v \right) \right] du\, dw \nonumber \\[.2cm]
       &+ v^2\, d\Omega^2. \nonumber 
\end{align}

By comparing the  coefficients of the terms $dw^2$ and $du\, dw$ between Eqs.~\eqref{eq:ds2Lake} and \eqref{eq:ds2LakeKottler}, one finds
\begin{subequations}
\begin{align}
&-f \left(\partial_w \tilde{t} \right)^2  + \frac{1}{f} \left(\partial_w v \right)^2=0 \, ,\\
&-f \left(\partial_u \tilde{t} \right)\left(\partial_w \tilde{t} \right) 
         + \frac{1}{f} \left(\partial_u v \right)\left(\partial_w v \right)=1 \, .
\end{align}
\end{subequations}
Therefore,
\begin{subequations}
\begin{align}
    \partial_w \tilde{t} & = \pm \frac{\partial_w v}{f} \, , \label{eq:partialwt}\\ 
    \partial_u\tilde{t}& = \pm \left(\frac{\partial_u v}{f} - \frac{1}{\partial_w v} \right)  \, , \label{eq:partialut}
\end{align}
\end{subequations}
where the $\pm$ sign in Eq.~\eqref{eq:partialut} is the same as Eq.~\eqref{eq:partialwt}. With the above, we conclude the map derivation between Lake and Kottler coordinates, which is given by
\begin{subequations}
\begin{align}
    \ell & = v(u,w)\, , \\[.2cm]
    \pm d\tilde t & = \left(\frac{\partial_u v}{f(v)} - \frac{1}{\partial_w v} \right)du + \frac{\partial_w v}{f(v)}  dw \, .
\end{align}
\end{subequations}

To find the correspondence with the SdS cosmological coordinates \eqref{eq:ds^2}, we use the map~\eqref{eq:CoordinateTransformations}, leading to the same map first shown in Eqs.~\eqref{eq:coord_map},
\begin{subequations} \label{eq:coord_map_app}
\begin{align}
    r a(t) & = v(u,w) \, , \\[.2cm]
    dt & =  \left[ \pm\left(1 - \frac{f(v)}{\partial_u v \partial_wv} \right) - \sqrt{\frac{\Lambda}{3}} \frac{v}{A(v)} \right] \frac{\partial_u v}{f(v)}du \nonumber \\
    & +\left( \pm 1 - \sqrt{\frac{\Lambda}{3}} \frac{v}{A(v)} \right) \frac{\partial_w v}{f(v)} dw \, .
\end{align}
\end{subequations}
In the above, the $\pm$ signs are correlated. For $\dot a < 0$, the signs in front of the $\sqrt{\Lambda/3}$ terms should be inverted. One can further expand the map, writing all the terms with explicit dependence on $u$ and $w$ and on the constants $C$ and $m$, however the final expression is quite large. A computational code for the map verification is provided in \cite{rodrigues_SdS_Zenodo}, which also includes a map version with all the $u$ and $w$ terms explicitly written. 

\bigskip

With the map above we can compare the time orientation between the proposed coordinates and those of Lake. 

Following Lake \cite{Lake:2005bf}, radial null geodesics satisfy either one of the conditions below,
\begin{subequations}
\begin{align}
    du &= 0, \label{eq:geoLake1} \\
    \frac{du}{dw} &= -\frac{2}{g}.\label{eq:geoLake2}
\end{align}
\end{subequations}
The geodesics are parametrized with the affine parameter $w$, which is such that \( dw \geq 0 \). The null geodesics with $du=0$ is simpler to be studied analytically and we focus on it here. For $du=0$, negative and positive $u$ lead respectively to  $\partial_w v < 0$ and $\partial_w v > 0$ (implying that light rays are either getting closer to the singularity at $v=0$, or getting farther away). 

To understand how the $du=0$ geodesics appear in the \((t, r)\) coordinate system, we consider Eq.~\eqref{eq:coord_map}  using the plus sign, thus yielding
\begin{align} 
    dt & = \left( 1 - \sqrt{\frac{\Lambda}{3}} \frac{v}{A} \right) \frac{\partial_w v}{f} dw  \nonumber \\
    & = u \frac{3m - C}{C^2} \frac{1}{f} 
    \left(1 - \sqrt{\frac{\Lambda}{3}}  \frac{v}{A} \right) dw \\
    & = u \frac{3m - C}{C^2} \frac{2}{f + 2 v \sqrt{\frac{\Lambda}{3}} + \sqrt{f^2 + 4 \frac{\Lambda}{3} v^2 }}  dw \, .\nonumber
\end{align}    
The above equation is sufficient to  show that there is no singularity in the map for $du=0$ geodesics and that the sign of $u$ has a direct impact on the relation between $dt$ and $dw$: for $u>0$, $dt$ has the same sign of $dw$; while for $u<0$ the situation is reversed (this is independent on the value of $f$). The case $du=0$ with $u<0$ is the situation in \cite{Lake:2005bf} that shows light coming from the cosmological horizon and entering the BH; but such case requires an inversion of the cosmological time with respect to the $w$ evolution. With the map \eqref{eq:coord_map} it is possible to find such geodesics with $dt>0$ and $dw>0$ in the region $u<0$, but this requires changing the map $\pm$ sign in front of 1 together with using cosmological contraction $\dot a<0$. 

We consider now the other geodesics \eqref{eq:geoLake2}. For this case, the term that multiplies $du$ in \eqref{eq:coord_map_app} is relevant, and $du$ can be replaced by $-2 \, dw /g$. This term also includes  the term $(1 - \sqrt{\Lambda/3} v/A)/f$, which was already shown to be regular at $f=0$, or any other positive value of $v(u,w)$. However, the term $g^{-1}$ introduces a singularity at $w=0$. And the term $(\partial_w v)^{-1} g^{-1}$ is singular both at $w=0$ and $u=0$. Hence, although both coordinate systems are regular at the horizons, their mutual map is not. This is consistent with the fact that, for fixed cosmological expansion, no geodesics in Fig.~\ref{fig:plotEtaRgeodesics} enter the BH  horizon.

\section{Ingoing and outgoing geodesics with respect to $\ell$} \label{app:HC}
Here we revisit Sec.~\ref{sec:causal} considering the ingoing and outgoing geodesics defined respectively by decresing and increasing $\ell$ (instead of $r$). 

Since $\ell = a r $, Eq.~\eqref{eq:dt2dlambda2} leads to
\begin{align}
    \frac{d \ell}{d\lambda} & = a \frac{dr}{d\lambda} + \ell \sqrt{\frac{\Lambda}{3}} \frac{dt}{d \lambda} \, . \nonumber \\[.2cm]
    &  = a \left|\frac{dr}{d\lambda}\right| \sqrt{\frac{\Lambda}{3}} \frac{\ell}{A}  \left(1 \pm \sqrt{\frac{3}{\Lambda}} \frac{A}{\ell} \right) \, .
\end{align}

We aim to understand the sign of $d\ell/d\lambda$, which only depends on the sign of the term inside parenthesis. For the positive sign, it is clear that $d\ell / d\lambda > 0$. For the negative sign, one has
\begin{equation}
    P \equiv 1 - \sqrt{\frac{3}{\Lambda}} \frac{A}{\ell} = 1 -  \tilde f(\ell) - \sqrt{1 + \tilde f^2(\ell)}  \, ,
\end{equation}
where $\tilde f(\ell) \equiv \sqrt{\frac{3}{\Lambda}} \, \frac{1}{2 \ell} \, f(\ell) $. One notes that $P$ can be written as a function of $\tilde f$, with $P(0) = 0$. Also, $\partial_{\tilde f}P < 0$. Therefore, $P < 0$ for $f > 0$ and $P > 0$ for $f<0$. In conclusion,
\begin{equation}
    \mbox{sign} \left(\frac{d\ell}{d\lambda}\right) = 
    \begin{cases}
        \mbox{sign}\left( \frac{dr}{d\lambda}\right) & \mbox{if} \quad \ell_1 < \ell < \ell_2 \\[.5cm]
        +1 & \mbox{if} \quad \ell < \ell_1 \quad \mbox{or} \quad \ell > \ell_2
    \end{cases} \, .
\end{equation}
This implies that only in the region between the horizons $\ell$ can both decrease or increase, in the other regions it can only increase. That is, the line element \eqref{eq:ds^2} can only cover the regions I, II and IV from Fig.~\ref{fig:penroseDiagram}.

\bibliography{KottlerCSF, general}

\end{document}